\documentclass[aps,prx,a4paper,twocolumn,longbibliography,floatfix]{revtex4-2}

\usepackage{amssymb,amsmath,amsfonts}
\usepackage{ dsfont }
\usepackage{comment}
\usepackage{amsxtra}
\usepackage{accents}
\usepackage{graphicx}
\usepackage[dvipsnames]{xcolor}
\usepackage[bookmarksopen,colorlinks,linkcolor=blue,citecolor=olive,urlcolor=red]{hyperref}
\usepackage{bm}

\DeclareMathOperator{\re}{Re}
\DeclareMathOperator{\im}{Im}
\DeclareMathOperator{\sign}{sign}
\DeclareMathOperator{\Tr}{Tr}

\def\EllipticE{E}

\def\br{\mathbf{r}}

\newcommand{\eps}{\varepsilon}
\newcommand{\corr}[1]{\langle#1\rangle}

\newcommand\smvee{\raise0.8ex\hbox{$\scriptscriptstyle\vee$}}

\def\be{\begin{equation}}
\def\ee{\end{equation}}

\begin{document}

\title{Plasmon excitations and their attenuation in dirty superconductors}

\author{Daniil K.\ Karuzin}
\affiliation{L.~D.\ Landau Institute for Theoretical Physics,  Chernogolovka 142432, Russia}
\affiliation{Moscow Institute of Physics and Technology, Dolgoprudny 141701, Russia}

\author{Mikhail A.\ Skvortsov}
\affiliation{L.~D.\ Landau Institute for Theoretical Physics,  Chernogolovka 142432, Russia}
\affiliation{Moscow Institute of Physics and Technology, Dolgoprudny 141701, Russia}

\date{\today}

\begin{abstract}
We develop a theory for the plasmon spectrum in dirty superconductors across the entire temperature range. Starting with the microscopic Keldysh sigma model description, we link the plasmon dispersion $\omega(q)$ to the optical conductivity $\sigma(\omega,T)$ of a superconductor, which requires analytical continuation to the lower half-plane of complex frequency. This approach reveals a discontinuity at the superconducting transition: a jump in both the real and imaginary parts of $\omega(q)$ at $T_c$. For any temperature below $T_c$, the plasmon dispersion terminates at a critical wave vector $q_c(T)$ where plasmons remains undamped, with 
weakly $T$-dependent 
$\omega[q_c(T)] \approx 2\Delta(0)$. 
Plasmons significantly attenuate only within a narrow 5\% 
window near $T_c$, with the propagating mode recovering at large~$q$.
\end{abstract}

\maketitle

\section{Introduction}
\label{S:Intro}

Collective excitations in superconductors have been the subject of intense theoretical and experimental study for decades \cite{Vaks,Schmid1968,CGexp,SS1975,Pethick,Mooij1985,Moor2017,Kulik1981,MillisCG,Shimano,Chubukov,Burmi2024CG,Burmi2024,DzKamenev2025}. The interplay between Coulomb interactions and fluctuations of the complex superconducting order parameter gives rise to a variety of subgap modes, such as plasmons \cite{Mooij1985}, the amplitude (Schmid-Higgs, SH) mode \cite{Schmid1968}, and the Carlson-Goldman (CG) mode \cite{CGexp}. The dispersion and attenuation of these collective modes are governed by nonequilibrium quasiparticle dynamics, and as such, are highly sensitive to the level of disorder.

The CG mode is characterized by synchronized oscillations of the normal and supercurrent, preventing the accumulation of space charge. 
This mode exists only in a close vicinity of the critical temperature, $T_c$, and exhibits a linear dispersion relation $\omega(q)$ \cite{CGexp,Mooij1985,Kulik1981,MillisCG,Burmi2024CG}. 
The recent revival of interest in the amplitude SH mode was triggered by its experimental observation using ultrafast terahertz (THz) spectroscopy \cite{Matsunaga2012,Matsunaga2013,Sherman2015}. The SH mode can be generated by electromagnetic radiation either through the application of a finite dc supercurrent \cite{Ovchinnikov1, Ovchinnikov2, Moor2017} or as a nonlinear response \cite{Silaev2019}.

The spectrum of plasmon excitation in low-dimensional superconductors is known in two limits. Near $T_c$, plasmons are underdamped only well below the gap \cite{Mooij1985,MillisCG}. At low temperatures, they propagate as charge oscillations in a distributed LC circuit formed by the kinetic inductance of the superconductor and its momentum-dependent effective capacitance. In superconducting nanowire singe photon detectors (SNSPDs) \cite{Goltsman01,Natarajan12,Zadeh2021}, low-temperature plasmons with a linear dispersion can be used to access a particular point through a delay time technique \cite{Berggren2016, Goltsman2024}.

In two-dimensional (2D) normal electron systems, experimental observation of the plasmon resonance \cite{Allen1977, Theis1980, Kuku2003} became possible through advances in fabricating high-mobility heterostructures. These experiments confirmed the theoretically predicted dispersion relation for a clean Fermi gas: $\omega_p^2(q) = n_\text{2D}q^2V_0(q)/m$,
where $n_\text{2D}$ is the 2D electron concentration, $V_0(q)$ is the Coulomb potential, and $m$ is the electron mass. Detecting narrow-linewidth plasmons requires extremely pure samples, where the momentum relaxation rate, $\tau^{-1}$, is very low. In the opposite limit of strong impurity scattering, $\omega_p(q)\tau\ll1$, the normal-metal plasmon becomes overdamped, with a purely relaxational dispersion
\be
\label{wn(q)}
  \omega(q) 
  =
  -i \gamma_n(q) .
\ee
The momentum-dependent attenuation rate is given by
\be
\label{gamma-def}
  \gamma_n(q)
  =
  2 \nu D q^2 V_0(q) ,
\ee
where $\nu$ is the density of states at the Fermi energy per one spin projection and $D$ is the diffusion coefficient.

Although dirty normal metals hold little interest for plasmonic applications, disorder is inherent to most materials used in contemporary superconducting electronics, such as SNSPDs and quantum phase-slip devices \cite{Mooij,Astafiev}.
Their core element is a very disordered superconducting film (NbN and similar compounds), approaching the metal-insulator transition.
This presents a challenge for systematically mapping the plasmon dispersion in dirty superconductors, particularly in understanding how the spectrum evolves from the overdamped regime above $T_c$ [Eq.\ \eqref{wn(q)}] to the non-decaying oscillations at absolute zero.

In this paper we develop a microscopic approach for studying the plasmon mode in dirty low-dimensional superconductors across the entire temperature range below $T_c$. Starting from the Keldysh sigma model for disordered superconductors \cite{SkvorKeld2000}, we derive the coupled equations of motion for the electric potential and the phase of the order parameter, which are bosonic degrees of freedom responsible for charge excitations. In the limit of small momenta, the plasmon dispersion $\omega(q)$ satisfies a nonlinear equation
\be
\label{spectral-eq-pl}
  \omega 
  = -i \frac{\sigma(\omega,T)}{\sigma_0} \gamma_n(q) ,
\ee
where $\sigma(\omega,T)$ is the complex optical conductivity of a superconductor \cite{MB}, and $\sigma_0 = 2 e^2 \nu D$ is the Drude conductivity in the normal state (we work in the system of units with $\hbar=1$). The spectral equation \eqref{spectral-eq-pl} defines a complex function $\omega(q)$ with the imaginary part $\im\omega(q)<0$ responsible for the plasmon attenuation.

\begin{figure}
\centering
\includegraphics[width=0.99\linewidth]{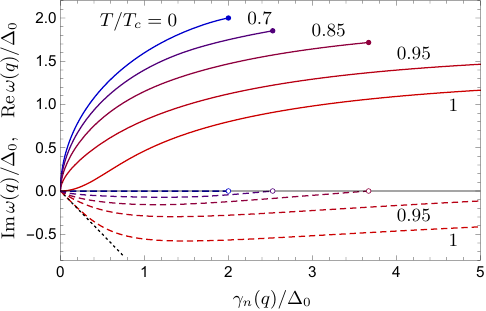}
\caption{Real (positive, solid) and imaginary (negative, dashed) parts of the plasmon dispersion $\omega(q)$ as a function of
$\gamma_n(q)/\Delta(0)$, for temperatures $T/T_c=0$, 0.7, 0.85, 0.95, and $1$ (just below the transition). For a given $T$, the spectrum terminates at a finite wave vector $q_c(T)$, where plasmon oscillations are undamped with frequency $\omega_c(T)$. In the normal state at $T>T_c$, plasmons are overdamped, $\omega(q)=-i\gamma_n(q)$ (dotted line in the lower left corner).}
\label{F:spectrum}
\end{figure}

While Eq.\ \eqref{spectral-eq-pl} has been derived previously under various approximations \cite{Mooij1985,MillisCG}, its rigorous solution requires the analytic continuation of the Mattis-Bardeen conductivity, $\sigma(\omega,T)$, into the lower half-plane $\im\omega<0$. To the best of our knowledge, this essential step has never been performed previously.
As a result, the plasmon dispersion was available only in the vicinity of $T_c$ and at low frequencies, $\omega\ll\Delta(T)$, where $\Delta(T)$ is the superconducting gap. 

The central difficulty in solving the spectral equation \eqref{spectral-eq-pl} lies in evaluating $\sigma(\omega,T)$ in the lower half-plane of complex $\omega$. This is non-trivial because the standard definition of the optical conductivity \cite{MB,fominov11,PolkSkv25} involves an integral over the real energy axis $\eps$, which becomes ill-defined upon analytic continuation. The reason is that for $\im\omega<0$, the branch cuts of the quasiparticle Green's functions cross the real $\eps$-axis. Thus, to analytically continue $\sigma(\omega,T)$ into the lower half-plane, one must perform a deformation the integration contour over $\eps$ to avoid these branch cuts.

One of the most striking consequences of the mentioned contour deformation is discontinuity of the plasmon dispersion $\omega(q)$ at the transition: both $\re\omega(q)$ and $\im\omega(q)$ exhibit jumps as $T$ is lowered below $T_c$. For a given momentum $q$, the attenuation rate, $-\im\omega(q)$, just below the transition (at $T_c-0$) is smaller than in the normal state, where it is given by $\gamma_n(q)$ [Eq.\ \eqref{gamma-def}]. At the same time, $\re\omega(q)$ appears abruptly below $T_c$. 

The plasmon dispersion, both $\re\omega(q)$ and $\im\omega(q)$, is presented in Fig.\ \ref{F:spectrum} for various temperatures, from absolute zero to $T_c$. For each temperature, the excitation spectrum has a termination point $q_c(T)$
(solutions of the spectral equation with $q>q_c(T)$ formally existing at the second Riemann sheet \cite{Chubukov,Burmi2024} do not correspond to well-defined physical excitations). 
At $T=0$, the undamped spectrum extends up to $\gamma_n[q_c(0)]=2\Delta(0)$, and the termination frequency $\omega_c(0)=2\Delta(0)$ coincides with the optical gap. In this case, the plasmon dispersion at $q\ll q_c(0)$ follows  the square-root law 
\be
\label{w0(q)}
  \omega_0(q) = \sqrt{\pi\Delta(0)\gamma_n(q)} .
\ee
This frequency corresponds to the resonance of a distributed LC circuit, where the inductance originates from the kinetic inertia of the Cooper pairs.

\begin{figure}
\centering
\includegraphics[width=0.99\linewidth]{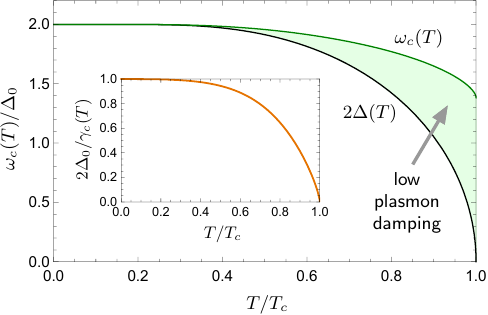}
\caption{Temperature dependence of the plasmon frequency at the spectrum termination point, $\omega_c(T)$ (upper curve).
At $T\ll T_c$, it goes exponentially close to $2\Delta(T)$ (lower curve), reaching the value of $1.382\,\Delta(0)$ at the transition.
Inset: momentum at the termination point, $q_c(T)$, quantified by 
$\gamma_c(T)=\gamma_n[q_c(T)]$.}
\label{F:termination}
\end{figure}

As temperature increases, the plasmon spectrum develops finite damping, shown by the dashed lines in Fig.~\ref{F:spectrum}. While the plasmon attenuation rate grows with $q$ in the long-wavelength limit, it reaches a maximum and subsequently decreases, falling to zero at the termination point, $q=q_c(T)$.

The behavior of the oscillation frequency at the spectrum termination point, $\omega_c(T)$, is shown in Fig.\ \ref{F:termination}.
Its value slightly exceeds the optical gap, $2\Delta(T)$, at low temperatures, yet it remains finite at $T_c$.
The existence of the region between $2\Delta(T)$ and $\omega_c(T)$ with suppressed plasmon damping is a direct manifestation of the dispersion jump at the critical temperature, which itself arises from the unique analytic structure of $\sigma(\omega)$ in the lower half-plane.
As a result, near $T_c$, plasmons which become damped once $\re\omega(q)$ reaches $\Delta^2(T)/T_c$ \cite{Mooij1985,MillisCG} can reappear as propagating excitations at larger wave vectors. This occurs when $\re\omega(q)$ enters the suppressed-damping region between $2\Delta(T)$ and $\omega_c(T)$.

The paper is organized as follows. In Sec.\ \ref{S:SE} we derive equations of motion for the electric potential and the order parameter phase, which are valid for any frequency, momentum and temperature. Analytical and numerical results for the plasmon dispersion are obtained in Sec.\ \ref{S:PD}. This Section also addresses the Carlson-Goldman mode for completeness. The results are discussed and summarized in Sec.\ \ref{S:DC}. Technical details of the sigma-model formalism are relegated to Appendix.

\section{Spectral equation}
\label{S:SE}

\subsection{Equation of motion for bosonic fields}

Collective excitations in a superconductor involve correlated dynamics of two bosonic fields: a complex order parameter $\Delta(\br,t)=|\Delta(\br,t)|e^{i\varphi(\br,t)}$ and a real electric potential $\phi(\br,t)$. In the absence of a dc supercurrent, the system's dynamics separate: the oscillations of the phase $\varphi(\br,t)$ and the potential $\phi(\br,t)$ are intrinsically coupled, while the amplitude (Schmid-Higgs) mode evolves independently \cite{Moor2017}. Thus, plasmon oscillations correspond to the combined dynamics of $\varphi(\br,t)$ and $\phi(\br,t)$ governed by the system of linear equations
\be
\label{eom}
\big(V^R\big)_{\omega,q}^{-1} 
  \begin{pmatrix}
    \phi_{\omega,q} \\ 
    |\Delta| \varphi_{\omega,q}
  \end{pmatrix} = 0 .
\ee
Isolating the contribution of the bare Coulomb interaction $V_0(q)$, we write $(V^R)^{-1}_{\omega,q}$ as
\be
\label{VR}
\big(V^R\big)^{-1}_{\omega,q}
  = 
  \begin{pmatrix}
    V_0^{-1}(q) & 0 \\
    0 & 0
    \end{pmatrix}
  + 2 \nu \Pi(\omega,q) ,
\ee
where $\Pi(\omega,q)$ is the retarded polarization matrix of the superconducting state. The latter is given by
\be
\label{Pi-def}
  \Pi(\omega,q)
  =
  \begin{pmatrix}
    1 & 0 \\
    0 & - 1/\lambda
  \end{pmatrix}
  +
  \frac{i}{4} \int d\eps \, \pi^R(\eps,\omega,q),
\ee
where the element 11 of the first (implicitly written) matrix describes static screening, $\lambda$ is the dimensionless BCS coupling constant, and the particular form of $\pi^R(\eps,\omega,q)$ explicitly depends on the level of disorder.

For a dirty diffusive superconductor, the matrix elements of $\pi^R(\eps,\omega,q)$ can be calculated straightforwardly by integrating out diffusive quasiparticle degrees of freedom within the Keldysh sigma model formalism \cite{SkvorKeld2000,SkvorKeld2018}, as outlined in Appendix \ref{A:Pi}. This procedure leads to the following expressions (hereafter we use a short-hand notation $\eps_\pm=\eps\pm\omega/2$):
\begin{subequations}
\label{pi's}
\begin{multline}
\label{pi_11}
  \pi^R_{11}
  = 
  (F_{\eps_+}-F_{\eps_-}) D_{\eps_+,\eps_-} \mathfrak{M}^{RA}_{\eps_+,\eps_-}
\\{}
  + 
  F_{\eps_-} C^R_{\eps_+,\eps_-} \mathfrak{M}^{RR}_{\eps_+,\eps_-}
  -
  F_{\eps_+} C^A_{\eps_+,\eps_-} \mathfrak{M}^{AA}_{\eps_+,\eps_-} ,
\end{multline}
\vspace{-20pt}
\begin{multline}
\label{pi_12}
  \pi^R_{12} = -\pi^R_{21} 
  = 
  (F_{\eps_+}-F_{\eps_-}) D_{\eps_+,\eps_-} \, \mathfrak{N}^{RA}_{\eps_+,\eps_-}
\\{}
  + F_{\eps_-} C^R_{\eps_+,\eps_-} \mathfrak{N}^{RR}_{\eps_+,\eps_-}
  - 
  F_{\eps_+} C^A_{\eps_+,\eps_-} \mathfrak{N}^{AA}_{\eps_+,\eps_-} ,
\end{multline}
\vspace{-20pt}
\begin{multline}
\label{pi_22}
  \pi^R_{22} 
  = 
  -
  (F_{\eps_+}-F_{\eps_-}) D_{\eps_+,\eps_-} \widetilde{\mathfrak{M}}^{RA}_{\eps_+,\eps_-}
\\{}
  - 
  F_{\eps_-} C^R_{\eps_+,\eps_-} \widetilde{\mathfrak{M}}^{RR}_{\eps_+,\eps_-}
  + 
  F_{\eps_+} C^A_{\eps_+,\eps_-} \widetilde{\mathfrak{M}}^{AA}_{\eps_+,\eps_-} .
\end{multline}
\end{subequations}
Here $F_\eps=\tanh(\eps/2T)$ is related to the equilibrium Fermi distribution function, 
$
  \mathfrak{M}^{\alpha_1\alpha_2}_{\eps_1\eps_2}
  =
  1 - g^{\alpha_1}_{\eps_1}g^{\alpha_2}_{\eps_2} - f^{\alpha_1}_{\eps_1}f^{\alpha_2}_{\eps_2}
$,
$
  \widetilde{\mathfrak{M}}^{\alpha_1\alpha_2}_{\eps_1\eps_2}
  =
  1 + g^{\alpha_1}_{\eps_1}g^{\alpha_2}_{\eps_2} + f^{\alpha_1}_{\eps_1}f^{\alpha_2}_{\eps_2}
$,
$
  \mathfrak{N}^{\alpha_1\alpha_2}_{\eps_1\eps_2}
  =
  g^{\alpha_1}_{\eps_1}f^{\alpha_2}_{\eps_2}-f^{\alpha_1}_{\eps_1}g^{\alpha_2}_{\eps_2}
$,
and the spectral functions $g_\eps$ and $f_\eps$ have the form
\be
\label{fg}
  \begin{pmatrix} 
    g^{R,A}_\eps \\ f^{R,A}_\eps
  \end{pmatrix}
  = 
  \frac{\pm1}{\sqrt{(\eps\pm i0)^2-\Delta^2}}
  \begin{pmatrix} 
    \eps\pm i0 \\ i\Delta
  \end{pmatrix}
  .
\ee
The diffuson and cooperon propagators on top of the superconducting state are given by
\begin{subequations}
\label{CD}
\begin{gather}
  D_{\eps,\eps'} 
  = 
  \frac{1}{D q^2 + \mathcal{E}^{R}_{\eps} + \mathcal{E}^{A}_{\eps'}},
\\{}
  C^\alpha_{\eps,\eps'} 
  = 
  \frac{1}{D q^2 + \mathcal{E}^{\alpha}_{\eps} + \mathcal{E}^{\alpha}_{\eps'}} ,
\end{gather}
\end{subequations}
where $\mathcal{E}^{R,A}_\eps = \mp i \sqrt{(\eps\pm i0)^2-\Delta^2}$.

Finally, the order parameter satisfies the self-consistency equation
\be
\label{sce}
  \frac{\Delta}{\lambda} 
  = 
  \frac{1}{2} \int d \eps \, F_\eps 
  \im 
  f^R_\eps .
\ee
Writing the term $1/\lambda$ in Eq.\ \eqref{Pi-def} with the help of Eq.~\eqref{sce} renders the energy integral for $\Pi_{22}(\omega,q)$ convergent in the ultraviolet.

\subsection{Plasmon spectral equation}

The dispersion relation $\omega(q)$ for collective excitations is obtained from the consistency condition for the system \eqref{eom} that reads
$\det \big(V^R\big)_{\omega(q),q}^{-1} = 0$.
Using Eq.\ \eqref{VR}, we write the spectral equation as
\be
\label{spectral-eq-0}
  [2\nu V_0(q)]^{-1} \Pi_{22}(\omega,q)
  + \det \Pi(\omega,q) = 0 .
\ee
In its full form, Eq.\ \eqref{spectral-eq-0} is analytically intractable for arbitrary $q$, and its solution requires numerical methods.
A significant simplification occurs in the limit of small momenta. 

At zero momentum, Eq.\ \eqref{spectral-eq-0} is satisfied automatically. Indeed, due to the long-range character of the Coulomb interaction [diverging $V_0(q)$ at $q\to0$], the first term vanishes.
The matrix $\Pi(\omega,0)$ evaluates to
\begin{equation}
\label{pizeroq}
    \Pi(\omega,0)
    = 
    f(\omega,T)
    \begin{pmatrix}
      1 & i \omega/2\Delta \\ -i \omega/2\Delta & (\omega/2\Delta)^2
    \end{pmatrix},
\end{equation}
where
\begin{equation}
\label{f-def}
    f(\omega,T) = \int_{\Delta}^\infty d \eps\frac{\Delta^2\tanh (\eps/2T)}{\sqrt{\eps^2-\Delta^2}
    [\eps^2 - (\omega/2 + i0)^2]} .
\end{equation}
Since $\det\Pi(\omega,0)=0$, Eq.\ \eqref{spectral-eq-0} holds true for arbitrary temperatures.
Such a behavior is a direct manifestation of electroneutrality. For spatially uniform oscillations, strong Coulomb interaction enforces $\phi = \dot\varphi/2$, which keeps the electrochemical potential constant and suppresses charge density generation.

Introducing the normal-state plasmon relaxation rate $\gamma_n(q)$ according to Eq.\ \eqref{gamma-def} and utilizing $\det\Pi(\omega,0)=0$, we can rewrite Eq.\ \eqref{spectral-eq-0} in the compact and most general form
\be
\label{spectral-eq-01}
  \frac{\Pi_{22}(\omega,q)}{\gamma_n(q)}
  + \frac{\det\Pi(\omega,q)}{Dq^2}
  = 0 ,
\ee
which contains both plasmon and neutral CG excitations.
Further analysis involves expanding Eq.\ \eqref{spectral-eq-01} in $Dq^2$, which produces multiple terms whose relevance depends on the specific mode.
The ratio of the relevant energy scales for the two modes can be estimated as $\gamma_n(q)/Dq^2 = 2\nu V_0(q)$, which for the bare 2D Coulomb interaction is of the order of $\kappa_2/q\gg1$, where $\kappa_2$ is the Thomas-Fermi screening momentum.
Since these modes are well separated in the momentum space, they can be studied independently.

Since we are primarily interested in plasmon excitations, we retain only the relevant terms in Eq.\ \eqref{spectral-eq-01} for now, postponing the discussion of the CG mode to Sec.\ \ref{SSS:CG}.
To this end, we replace $\Pi_{22}(\omega,q)$ by $\Pi_{22}(\omega,0)$ given by Eq.\ \eqref{spectral-eq-01}, and substitute $\det\Pi(\omega,q)/Dq^2$ by the leading contribution, $\det'\Pi(\omega,0)$, where prime stands for the derivative with respect to $Dq^2$.
Evaluating $\det'\Pi$ with the help of Eq.\ \eqref{pizeroq}, we find
\be
\label{detPi1}
  \det\nolimits'\Pi(\omega,0)
  = 
  f(\omega,T)
  \left[
  \frac{\omega^2}{4\Delta^2} \Pi_{11}' + \frac{i \omega}{\Delta} \Pi_{12}' + \Pi_{22}'
  \right] \! .
\ee
Remarkably, this weighted sum of $\partial\Pi_{ij}/\partial Dq^2$ at zero momentum is expressed in terms of the ac conductivity of a BCS superconductor, $\sigma(\omega,T)$, by the relation
\be
\label{det'Pi-res}
  \det\nolimits'\Pi(\omega,0)
  =
  \frac{i \omega f(\omega,T)}{4 \Delta^2} \frac{\sigma(\omega,T)}{\sigma_0} .
\ee
As a result, the spectral equation for the plasmon mode acquires an extremely simply looking form of Eq.\ \eqref{spectral-eq-pl}.

The optical conductivity of a superconductor has been calculated by Mattis and Bardeen \cite{MB}. We write it in the form suggested recently in Ref.\ \cite{PolkSkv25}:
\be
\label{sigma/sigma0}
  \frac{\sigma(\omega)}{\sigma_0}
  =
  1 + K^\text{RA}(\omega) + K^\text{R}(\omega) ,
\ee
where
\begin{subequations}
\label{KRA&KR}
\begin{equation}
\label{KRA}
K^\text{RA}(\omega)
=
-\frac{1}{4\omega}\int d\eps \,
  (F_{\eps_+} - F_{\eps_-})
  (1 + g^R_{\eps_+}g^A_{\eps_-} -  f^R_{\eps_+}f^A_{\eps_-}),
\end{equation}
\begin{equation}
\label{KR}
K^\text{R}(\omega)
= \frac{1}{2\omega}\int d\eps \, F_{\eps_-} (
1- g^R_{\eps_+}g^R_{\eps_-} + f^R_{\eps_+}f^R_{\eps_-}
).
\end{equation}
\end{subequations}
The representation given by Eqs.\ \eqref{sigma/sigma0} and \eqref{KRA&KR} is mathematically equivalent to that derived in Ref.\ \cite{fominov11}, but offers distinct advantages for numerical analysis.

\begin{figure}
\centering
\includegraphics{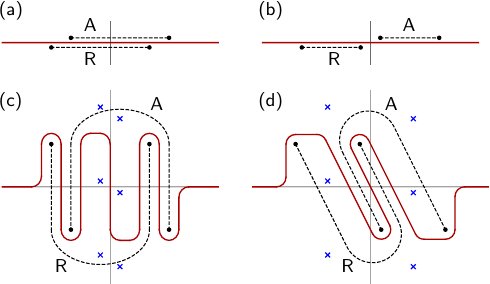}
\caption{Integration contours over $\eps$ for the conductivity contributions $K^{RA}(\omega)$ [Eq.\ \eqref{KRA}] for real $\omega$ (a,b) and in the lower half-plane, $\im\omega<0$ (c,d). Dots are the branching points $\pm\Delta\pm\omega/2$, and branch cuts are shown by dashed lines. 
Crosses are the poles of $F_{\eps_\pm}$. 
Panels (a) and (c) show the regime $\omega<\Delta$, while panels (b) and (d) correspond to $\omega>\Delta$.}
\label{F:contours}
\end{figure}

\subsection{Analytic continuation to $\im\omega<0$}
\label{SS:an-cont}

Since the optical conductivity is a complex quantity, the solution $\omega(q)$ for the spectral equation \eqref{spectral-eq-pl} is also complex, with a \emph{negative} imaginary part describing plasmon attenuation. This requires evaluating $\sigma(\omega,T)$ as a function of complex frequency $\omega$ in the lower half-plane. The representation \eqref{sigma/sigma0} is well suited for analytic continuation, which can be achieved by a proper deformation of the integration contour over $\eps$ to bypass the branching points of the Green functions \eqref{fg}.
For real $\omega$, they are located at $\pm\Delta\pm(\omega/2+i\delta)$, where the sign of an infinitesimal imaginary part $\delta$ is determined by causality. 
It is positive for $g^R$, $f^R$ and negative for $g^A$, $f^A$, ensuring that the branch cuts for the retarded and advanced Green's functions lie below and above the real axis, respectively, as shown in Fig.\ \ref{F:contours}(a,b).

When $\omega$ is shifted away from the real axis to the lower half-plane ($\im\omega<0$), one (for $K^{R}$) or two (for $K^{RA}$) branch cuts cross the real axis. Therefore the required contour deformation looks differently for the two terms in Eq.\ \eqref{sigma/sigma0}.
For the contribution $K^{R}(\omega)$, it suffices to draw the contour above $\pm\Delta-\omega/2$. In the case of $K^{RA}(\omega)$, the situation is more tricky as the branching points of the retarded ($\pm\Delta-\omega/2$) and advanced ($\pm\Delta+\omega/2$) Green functions move upward and downward, respectively. The proper choice of the contour providing analyticity of $\sigma(\omega)$ for $\im\omega<0$ is shown in Fig.\ \ref{F:contours}(c,d).

\section{Plasmon dispersion}
\label{S:PD}

\subsection{Zero temperature}

We start by analyzing the general expression \eqref{spectral-eq-pl} with the limit of zero temperature, when the optical conductivity is known exactly. The original Mattis-Bardeen paper \cite{MB} and the textbook \cite{Tinkham} give lengthy, separate expressions for $\re\sigma(\omega)$ and $\im\sigma(\omega)$. These can be simplified considerably by exploiting identities between elliptic integrals, leading to a concise form valid on the entire real frequency axis:
\be
\label{sigma-T0}
  \frac{\sigma(\omega,0)}{\sigma_0}
  =
  \frac{2i\Delta}{\omega} E(\omega/2\Delta) ,
\ee
where $E$ is the complete elliptic integral of the second kind, and $\Delta=\Delta(0)$.

Substituting Eq.\ \eqref{sigma-T0} into Eq.\ \eqref{spectral-eq-pl}, we arrive at the spectral equation at $T=0$:
\begin{equation}
\label{wzeroT}
  \omega^2 
  =
  2 \Delta \gamma_n(q) \EllipticE(\omega/2\Delta) 
  .
\end{equation}
As long as $\gamma_n(q)\leq2\Delta$, plasmons are undamped, with the dispersion relation shown in Fig.\ \ref{F:spectrum}. At lowest wave vectors, $\gamma_n(q)\ll2\Delta$, the elliptic integral may be replaced by $E(0)=\pi/2$, recovering the standard behavior \eqref{w0(q)}.

Since Eq.\ \eqref{wzeroT} comprises solely analytic functions, it can be directly evaluated for complex $\omega$. One can easily show that for $\gamma_n(q)>2\Delta$, this equation admits no solutions, real or complex. Hence, the wave vector $q_c(0)$, where $\gamma_n[q_c(0)]=2\Delta(0)$, is the spectrum termination point at zero temperature: No plasmon excitations---either damped or undamped---can propagate with $q>q_c(0)$. The corresponding spectrum termination frequency $\omega_c(0)=2\Delta(0)$, see Fig.\ \ref{F:termination}.

The elliptic integral $E(\omega/2\Delta)$ has two branch cuts: $(-\infty,-2\Delta)$ and $(2\Delta,\infty)$. One can construct a different function $E^\downarrow(z)$ that matches $E(z)$ in the upper-half plane, has a branch cut at $(-2\Delta,2\Delta)$, and thus is continuous at $|{\re \omega}| > 2 \Delta$. The function is defined as \cite{Chubukov}
\be
  E^\downarrow(z) 
  = 
  \begin{cases}
    E(z), & \im z > 0, \\ 
    E(z) - 2 i z [E(z) - K(z)/z^2], & \im z < 0.
  \end{cases}
\ee
Solution of the dispersion equation \eqref{wzeroT} with $E(z)$ replaced by $E^\downarrow(z)$ provides access to the second (nonphysical) Riemann sheet. Mathematically, at the termination point, the plasmon dispersion moves to the second sheet, saturating at $\omega = (2.309-0.857\,i) \Delta(0)$ at $q\gg q_c(0)$.

\subsection{Attenuation at low temperatures}

At finite temperatures, subgap plasmon modes attenuate due to excitations of thermal quasiparticles and the dispersion relation $\omega(q)$ acquires a negative imaginary part. In the low-temperatures limit ($T\ll T_c$), attenuation is weak and arises due to transitions between the states with $\eps\approx\Delta$ and $\eps + \omega$. Since the effect is exponentially small, we will focus only on imaginary (dissipative) terms in the spectral equation \eqref{wzeroT}. The low-$T$ corrections $\delta\sigma(\omega,T)=\sigma(\omega,T)-\sigma(\omega,0)$ can be evaluated from Eqs.\ \eqref{sigma/sigma0} by keeping the leading asymptotics of the spectral functions and diffusive modes at $\eps\to\Delta+\omega/2$, and taking into account an exponentially small deviation of $F_\eps$ from $\sign\eps$. Thereby we obtain
\be
  \frac{\delta\sigma(\omega,T)}{\sigma_0}
  =
  \frac{\Delta}{\omega}
  \sqrt{2 \pi\frac{T}{\omega} \frac{2 \Delta+\omega}{\Delta}}
  e^{-\Delta/T} \big[1 - e^{-\omega/T}\big] .
\ee

As a result, the right-hand side of Eq.\ \eqref{wzeroT} will acquire a small imaginary part $-i\omega [\delta\sigma(\omega,T)/\sigma_0] \gamma_n(q)$.
In the limit $\omega(q)\ll2\Delta$, we obtain
\be
  \omega(q) = \omega_0(q)
  - i \sqrt{\frac{T\omega_0(q)}{\pi}}
  e^{-\Delta/T} \big[1 - e^{-\omega_0(q)/T}\big] ,
\ee
where the first term is the zero-temperature plasmon dispersion \eqref{w0(q)}, while the second term describes attenuation due to quasiparticle excitation.

\subsection{Vicinity of $T_c$}

\subsubsection{Discontinuity at $T_c$}
\label{SS:discont}

In the normal state above $T_c$, the plasmon dispersion is given by the temperature-independent overdamped expression \eqref{wn(q)}.
Below $T_c$, a finite $\re\omega(q)$ emerges due to superconducting order parameter oscillations. Rather counterintuitively, $\omega(q)$ is discontinuous at the transition, with $\re\omega(q)$ appearing abruptly at $T<T_c$ and $|{\im\omega(q)}|$ jumping to a lower value (see Figs.\ \ref{F:spectrum} and \ref{F:w-vs-T}).

Such an unexpected behavior can be traced back to a peculiar counter-motion of the $g^R$ and $g^A$ branch cuts in the $K^\text{RA}(\omega)$ contribution to the conductivity when $\omega$ evolves to the lower half-plane, see Fig.\ \ref{F:contours}(c,d). In the limit of vanishing $\Delta$ but finite $\gamma$, the integration contour over $\eps$ is squeezed between the retarded and advanced branch cuts. Then we note that $g^R_{\eps_+}$ ($g^A_{\eps_-}$) takes its normal-state value 1 ($-1$) everywhere in the complex plane except for the interior $\mathfrak{S}^R$ ($\mathfrak{S}^A$) of a narrow strip formed by the retarded (advanced) branch cuts, where it equals $-1$ (1). Though these regions, where the Green functions differ by sign from their normal-state values, are infinitesimally narrow at $T\to T_c-0$, they are responsible for the discontinuity of the plasmon dispersion at $T_c$. The reason is that the integration contour has to follow a squeezed zigzag encompassing the branch cuts. Therefore, it is the length ($\propto\gamma$) rather than the width ($2\Delta$) of $\mathfrak{S}^{R,A}$ that is relevant. After discarding the term $f^R_{\eps_+}f^A_{\eps_-}$ proportional to $\Delta^2$, the integrand in Eq.\ \eqref{KRA} contains $1+g^R_{\eps_+}g^A_{\eps_-}$, which equals 2 at the symmetric difference of $\mathfrak{S}^R$ and $\mathfrak{S}^A$, and zero otherwise. Therefore we obtain
\begin{multline}
\label{KRA-at-Tc}
  K^\text{RA}(\omega)
  =
  -
  \lim_{\Delta\to0}
  \left( 
    \int_{-\omega/2}^{\omega/2-\Delta} 
    +
    \int_{\Delta-\omega/2}^{\omega/2} 
  \right) d\eps \,
  \frac{F_{\eps_+} - F_{\eps_-}}{2\omega}
\\{}
  =
  -
  \int_{-\omega/2}^{\omega/2} 
  d\eps \,
  \frac{F_{\eps_+} - F_{\eps_-}}{\omega}
  =
  - \frac{4T}{\omega} \ln\cosh \frac{\omega}{2T} .
\end{multline}
Note that this result derived in the limit $\Delta\to0$ is valid for any relation between $\omega$ (provided $\im\omega<0$) and $T_c$.

Expression \eqref{KRA-at-Tc} gives the jump in the plasmon dispersion at $T_c$. Just below the superconducting transition, at $T=T_c-0$, the spectrum satisfies
\be
\label{jump-eq}
  \omega 
  = - i \left( 1 - \frac{4T_c}{\omega} \ln\cosh \frac{\omega}{2T_c} \right) \gamma_n(q) .
\ee
This equation defines $\omega(q)/T_c$ as a function of $\gamma_n(q)/T_c$.
The plasmon dispersion calculated from Eq.\ \eqref{jump-eq} is shown in Fig.\ \ref{F:spectrum}, where it is plotted vs.\ $\gamma_n(q)/\Delta(0)$. 

In the limit of small $q$ and weak normal-state damping, $\gamma_n(q)\ll T_c$, plasmons at $T_c-0$ are strongly damped:
\begin{subequations}
\label{w(q)-discont-asymp}
\be
\label{w(q)-discont-1}
  \omega(q) = -i \gamma_n(q) + \frac{\gamma_n^2(q)}{2T_c}
  + i \frac{\gamma_n^3(q)}{4T_c^2} + \dots
\ee
As $q$ increases and eventually reaches the limit $\gamma_n(q)\gg T_c$, plasmons at $T_c-0$ become weakly damped:
\be
\label{w(q)-discont-2}
  \omega(q) 
  = 
  2.438 \, T_c - 8.756 i\, \frac{T_c^2}{\gamma_n(q)}
  - 46.21 \frac{T_c^3}{\gamma_n^2(q)}
  + \dots 
\ee
\end{subequations}
In the limit of large momenta, the plasmon dispersion just below the transition saturates at a finite frequency $\omega_c(T_c-0) = 2.438 \, T_c = 1.382\,\Delta(0)$, setting the lower bound for the spectrum termination frequency $\omega_c(T)$, as shown in Fig.\ \ref{F:termination}.

\begin{figure}
\centering
\includegraphics[width=0.99\linewidth]{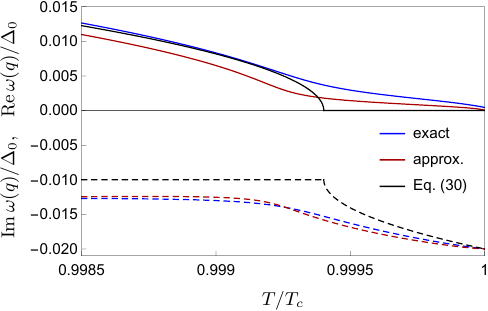}
\caption{Detailed view near $T_c$ of the temperature-dependent $\re\omega(q)$ (solid) and $\im\omega(q)$ (dashed) for $\gamma_n(q)/\Delta(0)=0.02$: exact behavior (blue), numerical solution based on the approximate conductivity \eqref{sigma-MB-GL} (red), and crude estimate \eqref{w0-near-tc} (black).}
\label{F:near-tc}
\end{figure}

\subsubsection{Behavior near $T_c$ at weak damping}
\label{SSS:Tc-weak-damping}

Here we discuss the temperature dependence of the dispersion relation $\omega(q)$ in the experimentally relevant weak-damping regime, $\gamma_n(q)\ll T_c$. Just below $T_c$, the real and imaginary parts of $\omega(q)$ given by Eq.~\eqref{w(q)-discont-1} satisfy the chain of inequalities
\be
   \re \omega(q) \ll |{\im\omega(q)}| \ll T_c .
\ee
Comparing these energy scales with the Ginzburg-Landau value of the order parameter, $\Delta(T)=\sqrt{8 \pi^2/7 \zeta(3)} \sqrt{T_c (T_c - T)}$, we see that $2\Delta(T)$ reaches $\re\omega$ at $T'=T_c(1-\tau')$ and $\im\omega$ at $T''=T_c(1-\tau'')$, where
\be
  \tau' 
  = 
  \frac{7\zeta(3)}{2^7\pi^2} \left( \frac{\gamma_n(q)}{T_c} \right)^4 ,
\quad
  \tau''
  = 
  \frac{7\zeta(3)}{2^5\pi^2} \left( \frac{\gamma_n(q)}{T_c} \right)^2 .
\ee
In the considered limit of weak damping, $\tau'\ll\tau''$, while $\tau''$ is as small as $\tau''\approx10^{-3}$ already at $\gamma_n(q)/T_c\approx 0.2$.

At temperatures below $T''$, the distance between the branching points of the Green functions, $2\Delta(T)$ exceeds $|\omega|$, and one can use the low-frequency ($\omega\ll\Delta$) asymptotics of the Mattis-Bardeen conductivity near $T_c$ \cite{Ovchinnikov1}:
\be
\label{sigma-MB-GL}
  \frac{\sigma(\omega,T)}{\sigma_0}
  =
  1 + \frac{\Delta}{2T} \ln \frac{8 \Delta}{e \omega}
  + \frac{i\pi \Delta^2}{2T\omega} .
\ee
This equation derived at real $\omega$ can be straightforwardly analytically continued to $\im\omega<0$ provided $|\omega|\ll\Delta$. With the conductivity \eqref{sigma-MB-GL}, the spectral equation \eqref{spectral-eq-pl} can be solved iteratively. First we neglect the logarithmic term and get an approximate solution ($\tau=1-T/T_c$)
\be
\label{w0-near-tc}
  \omega_0(q)
  = 
  - i\gamma_n(q) \frac{1+\sqrt{1-\tau/\tau_*}}{2} ,
\ee
where
\be
  \tau_*
  = 
  \frac{7\zeta(3)}{2^4\pi^2} \frac{\gamma_n(q)}{T_c}
\ee
is parametrically larger than $\tau'$ and $\tau''$ but is still much smaller than 1. 

The temperature dependence of the first approximation \eqref{w0-near-tc} for $\gamma_n(q)/\Delta(0)=0.02$ is shown in Fig.\ \ref{F:near-tc} (dotted lines), capturing the fully overdamped behavior above $T_*=(1-\tau_*)T_c$.
Though it qualitatively explains the main features of the true dispersion (solid lines), the effects of the neglected logarithm (dashed lines) become significant near $T_*$. These effects smear the square-root singularity in Eq.\ \eqref{w0-near-tc}, producing a finite $\re\omega(q)$ for all temperatures up to $T_c$.
Nevertheless, the plasmon remains overdamped (i.\ e., has a small quality factor) for $T>T_*$. The logarithmic term in the conductivity \eqref{sigma-MB-GL} is also responsible for a slightly visible maximum in $\im\omega(q)$ near $T_*$. 

The same analysis based on the spectral equation \eqref{spectral-eq-pl} and the conductivity \eqref{sigma-MB-GL} has been performed in Ref.~\cite{MillisCG}. Their Eq.\ (56), derived by neglecting the logarithmic term, is identical to our Eq. \eqref{w0-near-tc} and can be improved as discussed above.

What cannot be improved within the approach of Ref.\ \cite{MillisCG} is the lack of the dispersion jump at $T_c$. This jump only emerges when the $\omega\ll\Delta$ asymptotics \eqref{sigma-MB-GL} is superseded by the full analytic continuation of the conductivity to the region $-\im\omega\gg\Delta$, a key advancement described in Sec. \ref{SS:discont}.

\subsubsection{Carlson-Goldman mode near $T_c$}
\label{SSS:CG}

For pedagogical purposes, we demonstrate here how the CG mode can be obtained from the general spectral equation \eqref{spectral-eq-01}. This charge-neutral mode corresponds to the limit of an infinitely strong Coulomb interaction, which causes the first term in Eq.\ \eqref{spectral-eq-01} to vanish. Therefore, the remaining second should be expanded to the leading and subleading orders:
\be
\label{sp-eq-CG-1}
  \det\nolimits' \Pi(\omega,0)
  + \frac{Dq^2}{2} \det\nolimits'' \Pi(\omega,0)
  = 0 .
\ee
We now simplify this equation in the usual limit $\omega\ll\Delta$.
The first term in Eq.\ \eqref{sp-eq-CG-1} is given by Eq.\ \eqref{det'Pi-res}, with the conductivity provided by Eq.\ \eqref{sigma-MB-GL} and $f(\omega,T) \approx \pi \Delta/4T$.
The second term has two contributions, depending on how the derivatives are distributed between the matrix elements: $\det''\Pi(\omega,q) = \mathfrak{D}_1 + \mathfrak{D}_2$. Here $\mathfrak{D}_1 = 2 \det \Pi'(\omega,0)$, and $\mathfrak{D}_2$ arises when each matrix element in $\det\Pi$ is differentiated twice:
\be
\label{Pi21}
  \frac{\mathfrak{D}_2}{f(\omega,T)} 
  = 
  \frac{\omega^2}{4\Delta^2} \Pi_{11}'' 
  + \frac{i \omega}{\Delta} \Pi_{12}'' + \Pi_{22}''
  = 
  \frac{1}{2\Delta^2}
\ee
that holds for arbitrary $\omega$ and $T$. In the limit considered, $\mathfrak{D}_2 \approx \pi/8T\Delta$, while 
$\mathfrak{D}_1$ is determined by the low-frequency asymptotic behavior 
$\Pi_{11}'(\omega,0) = i/\omega + O(\omega)$, 
$\Pi_{12}'(\omega,0) = O(\omega)$, 
$\Pi_{22}'(\omega,0) = - \pi/8T + o(\omega\ln\omega)$. 
Therefore, $\mathfrak{D}_1=-i\pi/4T\omega$.
Substituting $\det''\Pi(\omega,q) \approx \mathfrak{D}_1$ into Eq.\ \eqref{sp-eq-CG-1} and omitting the small logarithmic term in Eq.\ \eqref{sigma-MB-GL}, we arrive at the spectral equation for the CG mode:
\be
\label{spectral-eq-2}
  \omega^2
  + \frac{i\pi\Delta^2}{2T} \omega
  - 2 D \Delta q^2 = 0 .
\ee

The spectral equation \eqref{spectral-eq-2} for the CG mode has been derived in Refs.\ \cite{Mooij1985,Kulik1981, MillisCG}. Its dispersion relation has the form
\be
\label{wCG}
  \omega_\text{CG}(q)
  =
  \sqrt{2 \Delta Dq^2 - \left(\frac{\pi \Delta^2}{4 T}\right)^2} 
  - i \frac{\pi \Delta^2}{4 T}.
\ee
In this approximation, the CG mode is overdamped at $q<q_0$, where $Dq_0^2=\pi^2\Delta^3/32T^2$. At large wave vectors, $q\gg q_0$, its dispersion becomes sound-like with the velocity $v=\sqrt{2 \Delta D}$ and small attenuation.

The above derivation of the CG mode's dispersion neglected the first, Coulomb term in Eq.\ \eqref{spectral-eq-01}. This is well justified for all practical purposes because the plasmon and CG modes are well separated in wave vector [a consequence of $\gamma_n(q)\gg Dq^2$] and the CG mode is a low-energy excitation.

\begin{figure}
\centering
\includegraphics[width=0.99\linewidth]{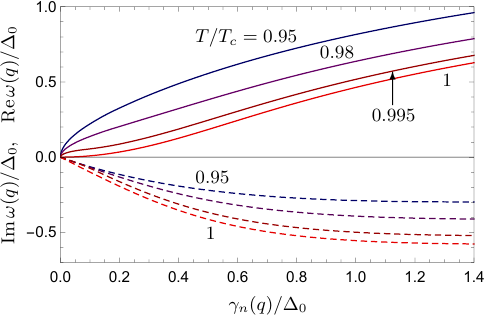}
\caption{Real (positive, solid) and imaginary (negative, dashed) parts of the plasmon dispersion $\omega(q)$ as a function of
$\gamma_n(q)/\Delta(0)$, for temperatures $T/T_c=0.95$, 0.98, 0.995, and $1$ (just below the transition). With increasing $T$, both $\re\omega(q)$ and $\im\omega(q)$ exhibit a downward shift.}
\label{F:ReIm}
\end{figure}

\subsection{Arbitrary temperatures}

At arbitrary temperatures, the plasmon dispersion $\omega(q)$ should be obtained numerically by solving Eq.\ \eqref{spectral-eq-pl}, where $\sigma(\omega)$ at $\im\omega<0$ must be calculated along a deformed contour to respect the analytical structure of the integrands in Eqs.\ \eqref{KRA&KR}, as explained in Sec.\ \ref{SS:an-cont}.

The complex plasmon dispersion, $\omega(q)$, simulated in this way is presented in Fig.\ \ref{F:spectrum} for temperatures $T/T_c=0$, 0.7, 0.85, 0.95, and $1-0$ (just below the transition). Its attenuation vanishes at absolute zero and increases towards $T_c$. Nevertheless, even at $T/T_c=0.95$, the plasmon is still weakly damped for all wave vectors.

A key feature of plasmon propagation in superconductors is the spectral termination at a critical wave vector $q_c(T)$, marked in Fig.\ \ref{F:spectrum}. For a given temperature, the plasmon dispersion is strictly limited to $q\leq q_c(T)$. Plasmons in the vicinity of the termination point are weakly damped, with their attenuation vanishing at $q_c(T)$. The oscillation frequency at the spectral termination, $\omega_c(T)=\omega[q_c(T)]$ is shown in Fig.~\ref{F:termination}. It decreases monotonically from $\omega_c(0)=2\Delta(0)$ to $\omega_c(T_c-0)=1.382\,\Delta(0)$, see Eq.\ \eqref{w(q)-discont-2}. The termination momentum $q_c(T)$ grows with increasing $T$ and diverges at the transition, see inset to Fig.\ \ref{F:termination}.
Matematically, at the termination point a plasmon leaves the physical Riemann sheet, reappearing on the second (unphysical) sheet. These hidden plasmon modes may manifest themselves indirectly through a peak in the response function \cite{Chubukov,Burmi2024}.

\begin{figure}
\centering
\includegraphics[width=0.99\linewidth]{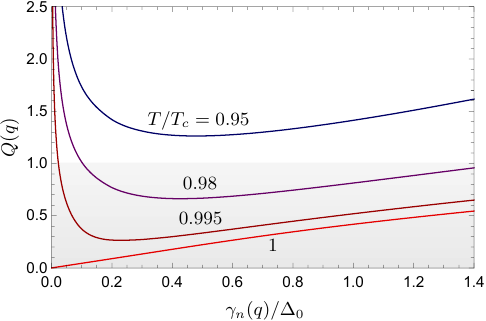}
\caption{Plasmon quality factor, $Q(q)=\re\omega(q)/2|{\im\omega(q)}|$, as a function of $\gamma_n(q)/T_c$, derived from the data in Fig.\ \ref{F:ReIm}.}
\label{F:Q}
\end{figure}

In the vicinity of $T_c$, the plasmon dispersion is illustrated in Fig.\ \ref{F:ReIm}, which shows $\omega(q)$ for several temperatures above $0.95\,T_c$. The corresponding quality factor, $Q=\re\omega(q)/2|{\im\omega(q)}|$, is plotted in Fig.\ \ref{F:Q}. The region near the transition provides an opportunity to benchmark our findings against earlier results \cite{Mooij1985,MillisCG}. At lowest $q$, the plasmon is a well-defined excitation with the predicted scaling $\omega(q)\propto\sqrt{\gamma_n(q)}$. According to Refs.\ \cite{Mooij1985,MillisCG}, it is expected to be damped when $\omega$ reaches $\Delta^2(T)/T_c$. Indeed, in Fig.\ \ref{F:Q} we see a strong reduction of the quality factor with increasing $q$. However, the reduction at low $q$ turns to an increase at larger $q$, producing a minimum in the momentum dependence of $Q$. This $Q_\text{min}$ is larger that 1 if $T<0.96\, T_c$. 

Consequently, a plasmon may become overdamped only within a very narrow temperature range, approximately $4\%$ of $T_c$. But even in this regime, increasing $q$ will transform a decaying plasmon into a propagating one. The observed revival of a plasmon at large $q$ is a consequence of the discontinuity at $T_c$, which leads to the growth of $\re\omega(q)$ and suppression of $|{\im\gamma(q)}|$ with increasing $q$.

The discontinuity at the transition is better visualized when the plasmon dispersion is plotted as a function of temperature at a given $q$, see Fig.\ \ref{F:w-vs-T}. A larger $q$ results in a higher starting value of $\re\omega(q)$ and a lower attenuation on the superconducting side of the transition.

\section{Discussion and conclusion}
\label{S:DC}

Collective excitations in a superconductor are governed by the quasiparticle response to time-dependent bosonic fields: the electric potential and the order parameter's phase and amplitude. This response is encoded in a $3\times3$ polarization operator, which incorporates the independent dynamics of the amplitude (SH) mode and a separate block describing coupled charge-phase oscillations. By neglecting the SH component, we derive an expression for the $2\times2$ matrix $\Pi(\omega,q)$ in a dirty superconductor, responsible for plasmons excitations and the CG mode.

\begin{figure}
\centering
\includegraphics[width=0.99\linewidth]{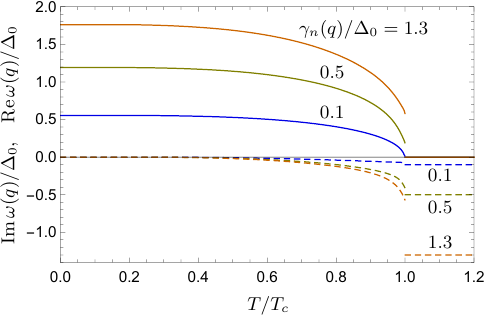}
\caption{Temperature dependence of the real (positive) and imaginary (negative) parts of the plasmon dispersion $\omega(q)$ calculated for three wave vectors corresponding to $\gamma_n(q)/\Delta(0) = 0.1$, 0.5 and 1.3. When the temperature drops below $T_c$, the normal-state overdamped dynamics $\omega(q)=-i\gamma_n(q)$ is abruptly modified, with an immediate appearance of a finite $\re\omega(q)$ and a jump in $\im\omega(q)$.}
\label{F:w-vs-T}
\end{figure}

Despite the generic complexity of the quasiparticle response, the optical conductivity $\sigma(\omega,T)$ is its only characteristic relevant for low-momentum plasmons. While the spectral equation \eqref{spectral-eq-pl} appears deceptively simple, solving it demands that $\sigma(\omega,T)$ be analytically continued to $\im\omega<0$, which is a challenging task. We achieve that by a proper deformation of the contour in the integral representation for the Mattis-Bardeen conductivity.

The peculiar structure of interpenetrating branch cuts of the retarded and advanced Green functions (Fig.\ \ref{F:contours}) has a number of important consequences:
\begin{itemize}
\item
The jump of the plasmon dispersion at $T_c$. While strong impurity scattering totally damps normal plasmons, a finite propagation frequency $\re\omega(q)$ immediately appears just below the superconducting transition. 
\item
Existence of the spectral termination point, $q_c(T)$, above which a plasmon does not exist as a physical excitation. Right at $q_c(T)$, a plasmon becomes undamped.
\item
Revival of propagating plasmons at large wave vectors near $T_c$.
\end{itemize}

\begin{figure}
\centering
\vspace{1mm}
\includegraphics[width=0.99\linewidth]{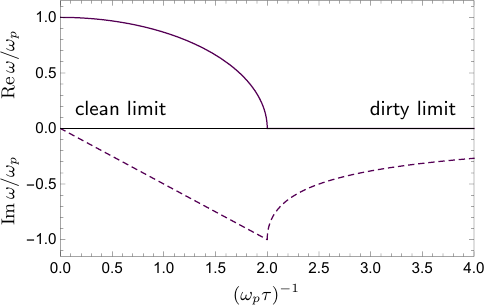}
\caption{Normal-metal plasmon dispersion $\omega(q)$ in the cross\-over between the clean and dirty limits controlled by the ratio of the momentum relaxation rate $\tau^{-1}$ and the clean plasmon frequency $\omega_p(q)$, according to Eq.\ \eqref{w0-metal}.}
\label{fig:ql}
\end{figure}

It is instructive to compare our results with the plasmon behavior in a normal metal film in the presence of momentum relaxation. 
The spectral equation follows from the vanishing of the dielectric function, which yields $V_0(q) P(\omega,q) + 1 = 0$, where $P(\omega,q)$ is the polarization operator (density-density correlation function).
Modelling impurity scattering by a Gaussian random potential with the correlator $\corr{U(\br)U(\br')} = \delta(\br-\br')/(2\pi\nu\tau)$ and calculating the impurity ladder \cite{Green}, we obtain
$P(\omega,q)/2\nu
  =
  1 - i\omega \tau [1-1/B(\omega,q)]^{-1}
$.
Given that plasmon wave vectors are typically very small, we evaluate the ladder block in the limit $ql\ll1$, while treating $\omega\tau$ as an arbitrary parameter: $1/B(\omega,q)=1 - i\omega \tau - Dq^2/(1 - i\omega \tau)$. The spectral equation then takes the form $i\omega(1-i\omega\tau)=\omega_p^2(q)\tau$, where $\omega_p(q)$ is the plasmon dispersion in the clean case: $\omega_p^2(q) = 2\nu D V_0(q)/\tau = \gamma_n(q)/\tau$. 
We thus arrive at the plasmon dispersion relation [mathematically equivalent to Eqs.\ \eqref{w0-near-tc} and \eqref{wCG}]
\be
\label{w0-metal}
  \omega(q)
  =
  \sqrt{\omega_p^2(q) - 1/4\tau^2} - i/2\tau .
\ee
For a given $q$, the evolution of $\omega(q)$ with the momentum relaxation rate $\tau^{-1}$ is presented in Fig.\ \ref{fig:ql}. By increasing the effective disorder strength, the system exhibits a crossover from the narrow plasmon resonance regime at $\tau^{-1}\ll1/\omega_p(q)$ to the purely overdamped regime at $\tau^{-1}>2/\omega_p(q)$. In the latter case, cooperative plasmon oscillations cannot be established due to dissipative losses, and their dispersion becomes purely relaxational [Eq.\ \eqref{wn(q)}], with the attenuation approaching $\gamma_n(q)$ given by Eq.\ \eqref{gamma-def}.

Comparing Figs.\ \ref{F:w-vs-T} and \ref{fig:ql}, one can qualitatively say that the transition to the superconducting state in a dirty metal effectively increases the apparent sample purity, restoring collective oscillations at low temperatures. The damping decreases with decreasing temperature, effectively mimicking the trend from the dirty to the clean case in Fig.\ \ref{fig:ql}. Of course, the clean plasma frequency, $\omega_p(q)$, is not recovered in the superconducting state. The reappearance of plasmon modes is caused not by the suppression of scattering, but rather by the emergence of a coherent, dissipationless motion in the superconducting condensate.

In dirty superconducting films, the plasmon wavelength is large. To estimate it, consider the $T=0$ spectral termination point determined by $\gamma_n[q_c(0)]=2\Delta(0)$. With $V_0(q) = 2 \pi e^2/\epsilon q$, one obtains for the typical wave vector: $q_c(0) = R_\Box \Delta(0) \epsilon/\pi$, where $R_\Box$ is the sheet resistance and $\epsilon$ is the effective dielectric constant. Assuming the BCS relation, $\Delta(0)=1.76\,T_c$, we obtain for the associate wave length
\be
\label{lambda-c}
  \lambda_c(0)
  =
  \frac{2\pi}{q_c(0)}
  =
  \frac{0.8\,\text{mm}}
  {R_\square[\text{k$\Omega$}] \, T_c[\text{K}] \,  \epsilon}  
  .
\ee
For a typical NbN film with $T_c \sim 10\,\text{K}$, $R_\Box \sim 2\,\text{k}\Omega$ and $\epsilon \sim 5$, we have $\lambda_c \sim 10\, \mu\text{m}$. This implies that the smallest structures capable of sustaining plasmonic resonance in dirty NbN films at $T\ll T_c$ must be on the scale of tens to hundreds of microns.

However, the rapid growth of the termination wave vector $q_c(T)$ near $T_c$ (see inset to Fig.\ \ref{F:termination}) shifts the experimental detection threshold for plasmon resonance to wavelengths smaller than those predicted by Eq.\ \eqref{lambda-c}. The experimental observation of weakly damped plasmons near $T_c$ is challenging, and would thereby offer crucial support for our theory.

\begin{acknowledgments}

We thank I. V. Kukushkin for initiating this research. This work was supported by the Russian Science Foundation under Grant No.\ 23-12-00297 and by the Foundation for the Advancement of Theoretical Physics and Mathematics ``BASIS''.

\end{acknowledgments}

\appendix

\section{Keldysh sigma model}
\label{A:Pi}

\subsection{Superconducting saddle point}

Nonequilibrium phenomena in dirty superconductors are most effectively described by the sigma model in the Keldysh formalism \cite{SkvorKeld2000}. It is a field theory for the quasiparticle field $Q(\br)$ subject to the constraint $Q^2=\openone$, the order parameter field $\mathbf{\Delta}(\br)$, and the electric potential field $\Phi(\br)$, specified by the action
\begin{multline}
\label{actionS}
  S
  =
  \frac{i \pi \nu}{4}
  \Tr \left[ D (\nabla Q)^2 + 4 i (i \tau_3 \partial_t + \check\Phi + i \check{\mathbf{\Delta}} )Q \right]
\\{}
  + \Tr \left[ \Phi^T (2\nu+V_0^{-1}) \sigma_1 \Phi 
  - (2\nu/\lambda) \mathbf{\Delta}^+ \sigma_1 \mathbf{\Delta} \right].
\end{multline}
The matrix $Q$ bearing two continuous energy/time indices acts in the direct product of the Gorkov-Nambu space (N, Pauli matrices $\tau_i$) 
and the Keldysh space (K, Pauli matrices $\sigma_i$). Depending on a single time/frequency argument, bosonic fields $\Phi = (\phi,\phi_q)^T$ and $\mathbf{\Delta}=(\Delta,\Delta_q)^T$ are two-component vectors in the Keldysh space, containing the classical and quantum components. They are organized in the $4\times4$ matrices
\begin{subequations}
\begin{gather}
  \check\Phi = \tau_0 (\phi \sigma_0 + \phi_q \sigma_1) ,
\\{}
  \check{\mathbf{\Delta}}
  =
  \begin{pmatrix}
        0 & \Delta \\ \Delta^* & 0
    \end{pmatrix}_\text{N} \sigma_0 + \begin{pmatrix}
        0 & \Delta_q \\ \Delta^*_q & 0
    \end{pmatrix}_\text{N} \sigma_1.  
\end{gather}
\end{subequations}

The stationary superconducting saddle point is characterized by the order parameter $\Delta$ (chosen to be real) and the 
energy-diagonal matrix $Q$ given by
\be
\label{Q-SC}
  Q(\eps)
  =
  \begin{pmatrix}
    Q^R(\eps) & F_\eps [Q^R(\eps)-Q^A(\eps)] \\ 0 & Q^A(\eps)
  \end{pmatrix}_\text{K},
\end{equation}
with the retarded ($\alpha=\text{R}$) and advanced ($\alpha=\text{A}$) blocks
\be
  Q^\alpha(\eps)
  =
  \begin{pmatrix}
    g^\alpha_\eps & f^\alpha_\eps \\ 
    f^\alpha_\eps & -g^\alpha_\eps
  \end{pmatrix}_\text{N}
\ee
composed of the Green functions introduced in Eq.\ \eqref{fg}.
At thermal equilibrium, $F_\eps = \tanh(\eps/2T)$.
The value of the order parameter should be obtained from the self-consistency equation \eqref{sce} [derivative of the action \eqref{actionS} with respect to $\Delta^*_q$].

\subsection{Diffusive modes}
 
Deviations from the superconducting saddle point \eqref{Q-SC} are parametrized by the matrix $W$ as \cite{SkvorKeld2000,SkvorKeld2018}
\be
\label{Q-W}
  Q = U^{-1}_F U^{-1} \sigma_3 \tau_3 (1 + W + W^2/2 + ...) U U_F .
\ee
The form of Eq.\ \eqref{Q-W} at $W=0$ is required to match Eq.\ \eqref{Q-SC}, which is enforced by the choice
\be
  U_F
  =
  \begin{pmatrix}
    1 & F_\eps \\ 0 & 1
  \end{pmatrix}_\text{K} , 
\quad 
  U
  =
  \begin{pmatrix}
    e^{i \tau_2 \theta^R_\eps/2}  & 0 \\ 0 & e^{i \tau_2 \theta^A_\eps/2}
  \end{pmatrix}_\text{K} ,
\ee
where the spectral angles $\theta^\text{R,A}_\eps$ 
parametrize the Green functions as $g^\text{R,A}_\eps = \pm\cos\theta^\text{R,A}_\eps$ and $f^\text{R,A}_\eps = \pm\sin\theta^\text{R,A}_\eps$.
The matrix $W$, which anticommutes with $\sigma_3 \tau_3$, has the form
\be
\label{difmodes}
     W = \begin{pmatrix}
        c_1^R i \tau_2 + c_2^R i \tau_1 & d_1 \tau_0 + d_2 \tau_3 \\ - \overline{d}_1 \tau_0 - \overline{d}_2 \tau_3 &   c_1^A i \tau_2 + c_2^A i \tau_1
    \end{pmatrix}_\text{K} .
\ee

At the Gaussian level, the correlation functions of these diffusive modes are given by
 \begin{gather}
     \langle d_{i;\eps_1,\eps_2}(\textbf{q}) \overline{d_i}_{\eps_3,\eps_4}(-\textbf{q})\rangle = \frac{1}{\pi \nu} D_{\eps_1,\eps_2}\delta_{\eps_1,\eps_4} \delta_{\eps_2,\eps_3},
\\{}
     \langle c^{R,A}_{i;\eps_1,\eps_2}(\textbf{q}) c^{R,A}_{i;\eps_3,\eps_4}(-\textbf{q})\rangle = \frac{1}{\pi \nu} C^{R,A}_{\eps_1,\eps_2}\delta_{\eps_1,\eps_4} \delta_{\eps_2,\eps_3} ,
 \end{gather}
where $\delta_{\eps,\eps'} = 2\pi \delta(\eps-\eps')$ and 
the bare diffuson and cooperon propagators are defined in
Eqs.\ \eqref{CD}.

\subsection{Effective action for bosonic fields}

Deviations of the bosonic fields from their equilibrium values are parametrized by $\delta \Phi = (\phi , \phi_q)^T$ and $\delta \mathbf{\Delta} = (\delta \Delta , \Delta_q)^T$.
Integrating out diffusive modes, one arrives at the effective action for the classical and quantum components of the fields $\delta\Phi$ and $\delta \mathbf{\Delta}$. In the problem of plasmon waves propagation, the electric potential is coupled only to the phase of the order parameter (in the linear order, $\varphi=\im \delta \Delta/\Delta$), while the amplitude Schmid-Higgs mode is not excited. (The latter will be admixed in the presence of a finite supercurrent \cite{Ovchinnikov2,Moor2017}.)
The resulting action takes the form
\be
  S^{(2)}_\text{eff}
  =
  \int (d \textbf{q})(d \omega) \, \eta^T_{-\omega,-q} \mathcal{V}_{\omega,q}^{-1} \, \eta_{\omega,q},
\ee
where we arrange the fields in a 4-component vector $\eta = (\phi, \im\delta\Delta, \phi_q, \im\Delta_q)^T$.

The inverse bosonic propagator $\mathcal{V}$ has the standard structure in the Keldysh space \cite{KA99,SkvorKeld2000}:
\be
\label{V4}
  \mathcal{V}_{\omega,q}^{-1}
  =
  \begin{pmatrix}
    0 & \big(V^A\big)_{\omega,q}^{-1} \\[3pt]
    \big(V^R\big)_{\omega,q}^{-1} & \big(V^K\big)_{\omega,q}^{-1}
  \end{pmatrix} .
\ee
Equations of motion for the classical components $\phi$ and $\im\Delta$ are obtained by taking the derivatives with respect to their quantum counterparts. This procedure cuts the block $(V^R)_{\omega,q}^{-1}$ from the matrix \eqref{V4}.
Calculating it in the RPA approximation, we arrive at Eqs.\ \eqref{eom}--\eqref{pi's}.


\begin{thebibliography}{99}

\bibitem{Vaks}
V. G. Vaks, V. M. Galitskii, and A. I. Larkin, 
Collective excitations in a superconductor, 
Zh. Eksp. Teor. Fiz. \textbf{41}, 1655 (1961)
[Sov. Phys. JETP \textbf{14}, 1177 (1962)].

\bibitem{Schmid1968}
A. Schmid, The approach to equilibrium in a pure superconductor: The relaxation of the Cooper pair density, 
Physik der kondensierten Materie \textbf{8}, 129 (1968).

\bibitem{CGexp}
R. V. Carlson and A. M. Goldman, 
Propagating order-parameter collective modes in superconducting films,
Phys. Rev. Lett. \textbf{34}, 11 (1975).

\bibitem{SS1975}
A. Schmid and G. Sch\"on,
Collective oscillations in a dirty superconductor,
Phys. Rev. Lett. \textbf{34}, 941 (1975).

\bibitem{Kulik1981}
I. O. Kulik, O. Entin-Wohlman, R. Orbach,
Pair susceptibility and mode propagation in superconductors: a microscopic approach,
J. Low Temp. Phys. \textbf{43}, 591 (1981).

\bibitem{Pethick}
C. J. Pethick and H. Smith, \emph{Nonequilibrium Superconductivity, Phonons and Kapitza Boundaries}, ed.\ by K. E. Gray (Plenum, New York, 1981).

\bibitem{Mooij1985}
J. E. Mooij and G. Sch\"on, Propagating plasma mode in thin superconducting filaments, Phys. Rev. Lett. \textbf{55}, 114 (1985).

\bibitem{Moor2017}
A. Moor, A. F. Volkov, and K. B. Efetov, Amplitude Higgs mode and admittance in superconductors with a moving condensate, Phys. Rev. Lett. 118, 047001 (2017).

\bibitem{MillisCG}
Z. Sun, M. M. Fogler, D. N. Basov, and A. J. Millis, 
Collective modes and terahertz near-field response of superconductors,
Phys. Rev. R \textbf{2}, 023413 (2020).

\bibitem{Shimano}
R. Shimano and N. Tsuji, 
Higgs mode in superconductors, 
Annu. Rev. Condens. Matter Phys. \textbf{11}, 103 (2020).

\bibitem{Chubukov}
D. Phan and A. V. Chubukov,
Following the Higgs mode across the BCS-BEC crossover in two dimensions,
Phys. Rev. B \textbf{107}, 134519 (2023).

\bibitem{Burmi2024CG}
E. S. Andriyakhina, P. A. Nosov, S. Raghu, and I. S. Burmistrov,
Quantum fluctuations and multifractally enhanced superconductivity in disordered thin films,
J. Low Temp. Phys. \textbf{217}, 187 (2024).

\bibitem{Burmi2024}
P. A. Nosov, E. S. Andriyakhina and I. S. Burmistrov, 
Spatially resolved dynamics of the amplitude Schmid-Higgs mode in disordered superconductors, 
Phys. Rev. B \textbf{111}, 144512 (2025).

\bibitem{DzKamenev2025}
M. Dzero and A. Kamenev, 
Schmid-Higgs mode in the presence of pair-breaking interactions,
Phys. Rev. B \textbf{111}, 174502 (2025).


\bibitem{Matsunaga2012}
R. Matsunaga and R. Shimano, 
Nonequilibrium BCS state dynamics induced by intense terahertz pulses in a superconducting NbN film,
Phys. Rev. Lett. \textbf{109}, 187002 (2012).

\bibitem{Matsunaga2013}
R. Matsunaga, Y. I. Hamada, K. Makise, Y. Uzawa, H. Terai, Z. Wang, and R. Shimano,
Higgs amplitude mode in the BCS superconductors Nb$_{1-x}$Ti$_x$N induced by terahertz pulse excitation,
Phys. Rev. Lett. \textbf{111}, 057002 (2013).

\bibitem{Sherman2015}
D. Sherman, U. S. Pracht, B. Gorshunov, S. Poran, J. Jesudasan, M. Chand, P. Raychaudhuri, M. Swanson, N. Trivedi, A. Auerbach, M. Scheffler, A. Frydman, and M. Dressel,
The Higgs mode in disordered superconductors close to a quantum phase transition,
Nat. Phys. \textbf{11}, 188 (2015).

\bibitem{Ovchinnikov1}
Yu. N. Ovchinnikov,
Properties of thin superconducting films in high frequency fields,
Zh. Eksp. Teor. Fiz. \textbf{59}, 128 (1970)
[Sov. Phys. JETP \textbf{32}, 72 (1971)].

\bibitem{Ovchinnikov2}
Yu. N. Ovchinnikov and A. R. Isaakyan,
Electromagnetic field absorption in superconducting films,
Zh. Eksp. Teor. Fiz. \textbf{74}, 178 (1978)
[Sov. Phys. JETP \textbf{41}, 91 (1978)].

\bibitem{Silaev2019}
M. Silaev,
Nonlinear electromagnetic response and Higgs-mode excitation in BCS superconductors with impurities,
Phys. Rev. B \textbf{99}, 224511 (2019).

\bibitem{Goltsman01}
G. N. Gol’tsman, O. Okunev, G. Chulkova, A. Lipatov, A. Semenov, K. Smirnov, B. Voronov, A. Dzardanov, C. Williams, and R. Sobolewski,
Picosecond super\-con\-duct\-ing single-photon optical detector,
Appl. Phys. Lett. \textbf{79}, 705 (2001).

\bibitem{Natarajan12}
C. M. Natarajan, M. G. Tanner, and R. H. Hadfield,
Superconducting nanowire single-photon detectors: physics and applications,
Supercond. Sci. Technol. \textbf{25}, 063001 (2012).

\bibitem{Zadeh2021}
I. E. Zadeh, J. Chang, J. W. N. Los, S. Gyger, A. W. Elshaari, S. Steinhauer, S. N. Dorenbos, V. Zwiller.
Superconducting nanowire single-photon detectors: A perspective on evolution, state-of-the-art, future developments, and applications,
Appl. Phys. Lett. \textbf{118}, 190502 (2021).

\bibitem{Berggren2016}
D. F. Santavicca, J. K. Adams, L. E. Grant, A. N. McCaughan, and K. K. Berggren,
Microwave dynamics of high aspect ratio superconducting nanowires studied using self-resonance,
J. Appl. Phys. \textbf{119}, 234302 (2016).

\bibitem{Goltsman2024}
V. Andreev, A. Semenov, N. Manova, V. Seleznev, S. Svyatodukh, A. Divochiy, P. Morozov, and G. Goltsman,
Dark counts in SNSPD studied with spatial resolution,
IEEE T. Appl. Supercon. \textbf{34}, 2200305 (2024).

\bibitem{Allen1977}
S. J. Allen, Jr., D. C. Tsui, and R. A. Logan,
Observation of the two-dimensional plasmon in silicon inversion layers,
Phys. Rev. Lett. \textbf{38}, 980 (1977).

\bibitem{Theis1980}
T. N. Theis,
Plasmons in inversion layers,
Surf. Sci. \textbf{98}, 515 (1980).

\bibitem{Kuku2003}
I. V. Kukushkin, J. H. Smet, S. A. Mikhailov, D. V. Kulakovskii, K. von Klitzing, and W. Wegscheider,
Observation of retardation effects in the spectrum of two-dimensional plasmons,
Phys. Rev. Lett. \textbf{90}, 156801 (2003).

\bibitem{Mooij}
J. E. Mooij and Y. V. Nazarov, 
Superconducting na\-no\-wi\-res as quantum phase-slip junctions,
Nat. Phys. \textbf{2}, 169 (2006).

\bibitem{Astafiev}
R. S. Shaikhaidarov, K. H. Kim, J. W. Dunstan, I. V. Antonov, S. Linzen, M. Ziegler, D. S. Golubev, V. N. Antonov, E. V. Il’ichev, and O. V. Astafiev,
Quantized current steps due to the a.c. coherent quantum phase-slip effect,
Nature \textbf{608}, 45 (2022).

\bibitem{SkvorKeld2000}
M. V. Feigel'man, A. I. Larkin, and M. A. Skvortsov, Keldysh action for disordered superconductors, Phys. Rev. B \textbf{61}, 12361 (2000).

\bibitem{MB}
D. C. Mattis and J. Bardeen,
Theory of the anomalous skin effect in normal and superconducting metals,
Phys. Rev. \textbf{111}, 412 (1958).

\bibitem{fominov11}
Ya. V. Fominov, M. Houzet, and L. I. Glazman, Surface impedance of superconductors with weak magnetic impurities, Phys. Rev. B \textbf{84}, 224517 (2011).

\bibitem{PolkSkv25}
A. V. Polkin and M. A. Skvortsov, Optical conductivity of a dirty current-carrying superconductor,  e-print arXiv:2512.06943.

\bibitem{SkvorKeld2018}
K. S. Tikhonov, M. A. Skvortsov, and T. M. Klapwijk, Superconductivity in the presence of microwaves: Full phase diagram, Phys. Rev. B \textbf{97}, 184516 (2018).

\bibitem{Tinkham}
M. Tinkham, \emph{Introduction to Superconductivity} (2nd ed., Dover Publications, New York, 2004).

\bibitem{Green}
L. S. Levitov and A. V. Shytov, \emph{Green Functions. Theory and Practice} (2nd ed., FizMatLit-Nauka, Moscow, 2007).

\bibitem{KA99}
A. Kamenev and A. Andreev, 
Electron-electron interactions in disordered metals: Keldysh formalism,
Phys. Rev. B \textbf{60}, 2218 (1999).

\end{thebibliography}
\end{document}